\documentclass{article}
\usepackage{amsmath}
\usepackage{tikz}
\usepackage{cite}
\usetikzlibrary{matrix}
\makeindex
\begin{document}
\title{\bf 3-Leibniz bialgebra in $N=6$ Chern-Simons gauge theories, multiple M2 to D2 branes and vice versa
 }
\author { M. Aali-Javanangrouh$^{a}$
\hspace{-2mm}{ \footnote{ e-mail:
aali@azaruniv.edu}}\hspace{1mm},\hspace{1mm} A.
Rezaei-Aghdam $^{a}$ \hspace{-2mm}{ \footnote{Corresponding author. e-mail:rezaei-a@azaruniv.edu }}\hspace{2mm}\\
{\small{\em
$^{a}$Department of Physics, Faculty of science, Azarbaijan Shahid Madani University }}\\
{\small{\em  53714-161, Tabriz, Iran  }}}
\maketitle
\begin{abstract}
Constructing  M2-brane and its boundary conditions from D2-brane and the related boundary conditions and vice versa has been possible in our recent work by using 3-Lie bialgebra for BLG model with $N=8$ supersymmetry. This could be generalized for BL model with $N=6$ by the concept of the 3-Leibniz bialgebra.  The 3-Lie bialgebra is an especial case of 3-Leibniz bialgebra, then more comprehensive information will be obtained in this work. Consequently, according to the correspondence of these 3-Leibniz bialgebras with Lie bialgebras, we reduce to D2-brane such that with some restrictions on the gauge field this D2-brane is related to the bosonic sector of an N = (4,4) WZW model equipped with one 2-cocycle in its Lie bialgebra structure. Moreover, the Basu-Harvey equation which is found by considering boundary conditions for BL model containing Leibniz bialgebra structure is reduced to Nahm equation and vice versa using this correspondence.

\end{abstract}

\section{\label{sec:intro}Introduction}
Unification of forces is one of the important issues in high energy physics. M-theory has been suggested for this purpose which can be obtained by developing the string theory \cite{Polchinski}. M-theory contains two types of branes, M2-brane and M5-brane \cite{Cremmer}. There are the great success \cite{Schmidhuber} in the description of action for single M2-brane. However, according to the existence of a self-dual field in the world volume of M5-brane and in spite of efforts, this theory is still unknown \cite{Howe}. It seems that some tricks may give us new information as the boundary conditions in the M2-brane ending to an M5-brane do \cite{Chu}.People attempted to find a Lagrangian in describing multiple M2-branes. Finally, Bagger and Lambert \cite{Bag1,Bag2,Bag3} and Gustavsson \cite{Gus} found one successful idea. Although, there was just one example \cite{Bag3} only for two membranes \cite{Raamsdonk}, the case has been considered by many authors \cite{Nilsson}.Bagger and Lambert inspired by the Basu-Harvey article \cite{Bas} (see also Ref.\cite{sheikh}) and groping supersymmetry transformations of D2-brane \cite{Witten, Mem} expressed $N=8$ supersymmetry transformations of M2-branes. They obtained a series of equations of motion according to supersymmetric transformations to be closure. They could be derived from a Lagrangian which Bagger and Lambert found it. It was interesting with some limitations as mentioned, so they attempted to find the other one for explaining multiple M2-brane. Using complex fields, Aharony, Bergman, Jafferis, and Maldacena (ABJM) \cite{ABJM} found a Lagrangian for the arbitrary number of membranes and reduced the supersymmetry to $N=6$. Bagger and Lambert followed this idea and described their own theory using complex scalar fields for arbitrary number of membranes \cite{Bag4}. In this way, they have got to use structure constants of 3-Lie algebra that is not totally antisymmetric. The relationship between those works has been described in Ref. \cite{Papageorgakis}.

The world volume of the membrane in M-theory and the equivalent one in string theory (D-brane) are in relation to each other by using the Higgs mechanism in $N=8$ BLG model \cite{Matsu,Mat}. The model and relations can be extended to $N=6$ BL model \cite{EMP} with some difference, such as attributing vacuum expectation value (VEV) to a real part of one of the complex scalar fields \cite{EMP} and also, the fialds of this model are 3-Leibniz algebra valued instead of 3-Lie algebra one. Note that, the structure constants in 3-Leibniz algebra are antisymmetric only in two indices, 3-Lie algebra will obtain from it by completely antisymmetric structure constants then the results of this work will involve the ones in \cite{ali}. On the other hand, 3-Lie algebra can be considered as a special case of 3-Leibniz algebra.  $N=6$ BL mode must have done with gauge fields in U(N) group instead of SU(N). One of the results in studying boundary condition for Lorentzian BL model  \cite{Sezgin,Passirini,Low,Lee} is   Basu-Harvey equation \cite{Bas} (as a boundary condition for the M2-brane ending to M5-brane) which will be turned into Nahm-equation \cite{Nahm} (as a boundary condition for D1-string ending D3-brane).  This work is not a simple generalization of $N=6$ \cite{Passirini}. It seems to be impossible to do the same in the inverse direction, but in this work, correspondence between 3-Leibniz
bialgebra  \cite{GR1} and Lie bialgebra \cite{Sch} we will show the possibility of it.

Blokh suggested the notion of Leibniz algebra  \cite{Blokh} in 1965; it was rediscovered by J. L. Loday \cite{Loday} with a non-antisymmetric bilinear bracket. Leibniz bialgebra \cite{GR} and 3-Leibniz bialgebra \cite{GR1} have been introduced recently by using cohomology of Leibniz and 3-Leibniz algebra.  We think that the role of 3-Leibniz bialgebra can be important in M-theory if we express $N = 6$ BL model in the 3-Leibniz bialgebra structure as an example. We have shown that there is a relation between these models and WZW models\footnote{One of the main descriptions to D-branes has been stated in terms of conformal field theory attributed to sigma-models on the world-sheet in Ref. \cite{Liegh}. D-branes on group manifolds can be described in terms of boundary states in a WZW model \cite{Kato}.} with Lie bialgebra structure equipped with one 2-cocycle \cite{AR}, this work with some difference has been done in \cite{Smith,Okazaki,Moore}. In the previous work \cite{Aali1} we have shown there is a relation between Nahm and Basu-Harvey equations and vice versa by considering their boundary conditions. In the other one  \cite{ali} we have shown that knowledge of the WZW model and its algebraic structure able one in increasing information about the $N=8$ BLG model and its algebraic structure, i.e., one will be able to construct a  relation between string theory and M-theory directly using 3-Lie bialgebra \cite{ali}. In this work using methods applied in Ref.\cite{ali}, we will try to extend the results to $N=6$ BL model by using the 3-Leibniz bialgebra \cite{GR1}. In this case, our results will be useful since it makes a relation between multiple M2-brane and D2-brane. The relation between Nahm and Basu- Harvey equations has been found by using the relation between 3-Lie algebra and super-Lie algebra \cite{Bielawski,Medeiros}.

The outline of this paper is as follows.  We review $N=6$ BL model in section 2 and consider boundary conditions for it in section 4. We will review and give a special example the 3-Leibniz bialgebra and show the relation between 3-Leibniz bialgebra and Lie bialgebra in section 3.  The BL model by 3-Leibniz bialgebra has been constructed as the works in Ref. \cite{ali,Aali1}. One of the most results is that  BL model has been turned into WZW model with Lie bialgebra structure which has been showed in section 4.  Also,  boundary conditions of the theory have been mentioned in section 5.
 
\section{\label{sec:N=6 BLG}N=6 BL model}
Here for self-containing of the paper, we review the $N=6$ BL model \cite{Bag4}. Dynamics of M2-brane in M-theory has been studied by maximal supersymmetric 3-dimensional Chern-Simons model \cite{Bag1,Bag2,Bag3,Gus}. Bagger and Lambert extended their model to arbitrary number of membranes with  $N= 6$ \cite{Bag4} supersymmetry using {\em special 3-Lie algebras}\footnote{We introduced it  Leibniz algebra in Ref. \cite{GR}} and complex scalar fields which have been deduced from the ABJM model \cite{ABJM}. In addition to the fundamental difference between ABJM and BL theories,  there is a conventional relation between them \cite{Papageorgakis}.The BL formulation has got four complex scalar fields $Z^{\alpha}$  (8 real field). Note that in  ABJM model  complex scalar fields  are split  into two pairs $A_a$ and ${\bar B}_a$ with $a=1,2$,   where $(Z^1 = A_1, Z^2 = A_2, Z^3 = \bar B_1, Z^4 = \bar B_2)$ which must be considered complex conjugates of each other ${\bar Z}_{\alpha{\bar a}}=(Z^{\alpha}_a)^*$  because of turning  SO(8) R-symmetry into $SU(4) \times U(1)$ \cite{Bag4}. One must have for  the fermion fields and their complex conjugates $\psi_{\alpha a}, \psi_{\bar a}^{\alpha}=(\psi_{\alpha a})^*$  where the  raised indice express  as
the fields  in the {\bf 4} of SU(4) and  a lowered indice convert in the ${\bf \bar 4}$. An index is raised or lowered by complex conjugation and supersymmetry generators $\epsilon_{\alpha \beta}$  that are in the {\bf 6} of SU(4) with zero charge of U(1)  and must satisfy the  relation as $\epsilon^{\alpha \beta}=\frac{1}{2}\varepsilon^{\alpha \beta \gamma \delta}\epsilon_{\gamma \delta}$ \cite{Bag4}. The  covariant derivative is
$D_\mu Z^{\alpha}_d = \partial_\mu Z^{\alpha}_d - \tilde{A} {{_\mu}^c}_d Z^{\alpha}_c, D_\mu \bar{Z}_{\alpha d} = \partial_\mu \bar{Z}_{ \alpha d} + \tilde{A} {{_\mu}^c}_d \bar{Z}_{\alpha c}\ 
$ where ${\tilde A}_{\mu d}^c = {f^{cb{\bar a}}}_d A_{\mu {\bar a}b}$ such that  $\alpha, \beta, \gamma \textrm{ and } \delta=1,2,3,4$ the numbers of the scalar fields, $\mu , \nu=0,1,2$ for representation of M2-brane wroldvolume, and the supersymmetry transformations are as follows:
\begin{eqnarray}
\delta Z^{\alpha}_d &=& i\bar\epsilon^{\alpha \beta}\psi_{\beta d} \\
\delta \psi_{\beta d} &=& \gamma^{\mu} D_{\mu} Z^{\alpha}_d\epsilon_{\alpha \beta} +
  f^{ab{\bar c}}_d Z^{\gamma}_a Z^{\alpha}_b {\bar Z}_{\gamma{\bar c}}\epsilon_{\alpha \beta}+
  f^{ab{\bar c}}_d Z^{\gamma}_a Z^{\delta}_{b} {\bar Z}_{\beta{\bar c}}\epsilon_{\gamma \delta} \\
\delta \tilde A_{\mu d}^c &=&
-i\bar{\epsilon}_{\alpha \beta}\gamma_{\mu} Z^{\alpha}_a\psi^{\beta}_{\bar b} f^{ca{\bar b}}_d +
i\bar{\epsilon}^{\alpha \beta}\gamma_{\mu} {\bar Z}_{\alpha{\bar b}}\psi_{\beta a} f^{ca{\bar b}}_d ,
\end{eqnarray}
with gauge transformations as:
\begin{equation}
\delta_{\Lambda}Z^{\alpha}_d=\Lambda_{{\bar c}b} f^{ab{\bar c}}_d Z_a^{\alpha}
\end{equation}
where can be showed by $
\delta_{\Lambda} Z^{\alpha}=N Z^{\alpha}-Z^{\alpha} N
$ such that 
$N$ and $M$ is $n\times n$ and  $m\times m$  matrix respectively, i.e., gauge group is $U(n)\times U(m)$ (see details in Ref. \cite{Bag4})\footnote{Note that the 3- Lie algebra $\cal A$ as vector space is a subspace of u(n) \cite{Bag4}.}. Closure of the supersymmetric  transformations  terminate the following equations of motion:
\begin{eqnarray}
0 &=& \gamma^\mu D_\mu \psi_{\gamma d} + {f^{abc}}_d\psi_{\gamma a}Z^{\delta}_b\bar{Z}_{\delta c} - 2{f^{abc}}_d\psi_{\delta a}Z^{\delta}_b\bar{Z}_{\gamma c} - \epsilon_{\gamma \delta \alpha \beta}{f^{abc}}_d\psi^{\delta}_cZ^{\alpha}_aZ^{\beta}_b\label{Psieom} \label{64}\\
0 &=& \tilde{F} {{_{\mu \nu}}^c}_d + \varepsilon_{\mu \nu \lambda} (D^\lambda Z^{\alpha}_a)\bar{Z}_{\alpha b}- Z^{\alpha}_a(D^\lambda\bar{Z}_{\alpha b}) - i\bar{\psi}^{\alpha}_b\gamma^\lambda\psi_{\alpha a}{f^{cab}}_d\ ,
\end{eqnarray}
and  equation of motion  of the scalar fields  achieved by  taking the supersymetric variation of (\ref{64}).
The invariant Lagrangian  related to this equations  can be expressed as follows \cite{Bag4}:
\begin{eqnarray}
\label{N6Bl}
\nonumber
 {\cal L} &=& - D^\mu {\bar Z}_{\alpha}^a D_{\mu} Z^{\alpha}_a -
i{\bar\psi}^{\alpha a}\gamma^{\mu} D_{\mu}\psi_{\alpha a} -V+{\cal L}_{CS}
\\
\nonumber
&-& i f^{ab{\bar c}{\bar d}}{\bar\psi}^{\alpha}_{\bar d} \psi_{\alpha a}
Z^{\beta}_b{\bar Z}_{\beta{\bar c}}+2if^{ab{\bar c}{\bar d}}{\bar\psi}^{\alpha}_{\bar d}\psi_{\beta a}Z^{\beta}_b{\bar Z}_{\alpha{\bar c}}
\\
&+&\frac{i}{2}\varepsilon_{\alpha \beta \gamma \delta}f^{ab{\bar c}{\bar d}}{\bar\psi}^{\alpha}_{\bar d}\psi^{\beta}_{\bar c} Z^{\gamma}_aZ^{\delta}_b
-\frac{i}{2}\varepsilon^{\alpha \beta \gamma \delta}f^{cd{\bar a}{\bar b}}{\bar\psi}_{\alpha c}\psi_{\beta d}{\bar Z}_{\gamma {\bar a}}{\bar Z}_{\delta{\bar b}}\
\end{eqnarray}
with
\begin{equation}
V = \frac{2}{3}\Upsilon^{\gamma \delta}_{\beta d}\bar\Upsilon_{\gamma \delta}^{\beta d} ,
\end{equation}
where
\begin{equation}
\Upsilon^{\gamma \delta}_{\beta d} = f^{ab{\bar c}}_dZ^{\gamma}_aZ^{\delta}_b{\bar Z}_{\beta{\bar c}}
-\frac{1}{2}\delta^{\gamma}_{\beta}f^{ab{\bar c}}_dZ^{\alpha}_aZ^{\delta}_b{\bar Z}_{\alpha \bar c}+\frac{1}{2}\delta^{\delta}_{\beta}f^{ab{\bar c}}_dZ^{\alpha}_aZ^{\gamma}_b{\bar Z}_{\alpha{\bar c}}.
\end{equation}
and Chern-Simons term have the following form:
\begin{equation}
\label{CSC}
{\cal
L}_{CS}=\frac{1}{2}\varepsilon^{\mu\nu\lambda}\left(f^{ab{\bar c}{\bar d}}A_{\mu{
\bar c} b}\partial_\nu A_{\lambda {\bar d} a} +\frac{2}{3}f^{ac{\bar d}}_gf^{ge{\bar b}f{\bar b}}
A_{\mu {\bar b} a}A_{\nu {\bar d} c}A_{\lambda {\bar b}f e}\right),
\end{equation}
where $f^{ac{\bar d}}_g$ is the structure constants of the 3-Leibniz algebra $\cal A$ with the following nonantisymmetric 3-bracket\footnote{$f^{ab{\bar c}{\bar d}}=-f^{ba{\bar c}{\bar d}}$ and $f^{ab{\bar c}{\bar d}}=f^{{\bar c}{\bar d}ab}$.}:
\begin{eqnarray}
    [T^a,T^b;T^{\bar c}]=f^{ab{\bar c}}\hspace{0cm}_d T^d,\hspace{1cm}a,b,c,d=1,...,dim{\cal A},
  \label{3-bracket}
\end{eqnarray}
such that we have  the following fundamental identity (see for example \cite{Farrill}):
\begin{eqnarray}
[[T^a,T^b;T^{\bar c}],T^d;T^{\bar e}]=[[T^a,T^d;T^{\bar e}],T^b;T^{\bar c}]+[T^a,[T^b,T^d;T^{\bar e}];T^{\bar c}]+[T^a,T^b;[T^{\bar c},T^d;T^{\bar e}]],
\label{fundamental identity}
\end{eqnarray}
where it can be redefined  by  the  structure constant of  ${\cal A}$ in the following form:
\begin{eqnarray}
f^{ab{\bar c}}\hspace{0cm}_ff^{fd{\bar e}}\hspace{0cm}_g- f^{ad{\bar e}}\hspace{0cm}_ff^{fb{\bar c}}\hspace{0cm}_g-f^{bd{\bar e}}\hspace{0cm}_ff^{af{\bar c}}\hspace{0cm}_g-f^{{\bar c}d{\bar e}}\hspace{0cm}_ff^{abf}\hspace{0cm}_g=0.
\end{eqnarray}

\section{\label{sec:3-Lie bialgebra1}3-Leibniz bialgebra}
In this section, we review the definition of 3-Leibniz bialgebra that have been  given recently in  \cite{GR1}.
\par
{\bf Definition}:  \cite{Sch} Lie bialgebra deals  with a map  $\delta : { \cal G} \rightarrow
{ \cal G}\otimes{ \cal G} $ such that:
\\
{\it a} ) $\delta$ is a one-cocycle, i.e.:
\begin{eqnarray}
\delta([T^{i},T^{j}])&=&{ad^{(2)}}_{T^j}\delta(T^{i})-{ad^{(2)}}_{T^i}\delta(T^{j}),
\end{eqnarray}
where
\begin{eqnarray}
{ad^{(2)}}_{T^j}&=&ad_{T^j}\otimes 1+1\otimes ad_{T^j},
\end{eqnarray}
 ${ \cal G}$ is a Lie algebra and $\{T^i\}$s are  bases for it,
\\
{\it b} ) $ ^t\delta:{ \cal G}^*\wedge{ \cal G}^* \rightarrow  { \cal G}^*$ operate as a cocommutator on dual space   of  ${ \cal G}$ i.e.  ${ \cal G}^*$ with the following relation:
 \begin{eqnarray}
\label{pairing}
(\tilde{T}_{i}\wedge \tilde{T}_{j},\delta(T^k))=({^t\delta(\tilde{T}_{i} \wedge \tilde{T}_{j}),T^k)=([\tilde{T}_{i},\tilde{T}_{j}]},T^k),
\end{eqnarray}
such that $ \{{\tilde T}_i\} $ are bases of ${\cal G}^*$.  ${ \cal G}^*$ is a Lie algebra and there is a pairing  between ${ \cal G}$ and ${ \cal G}^*$ shown as $(,)$. 

One can obtain following identities for Lie bialgebra:
\begin{eqnarray}
f^{ij}\hspace{0cm}_kf^{kl}\hspace{0cm}_m-f^{ik}\hspace{0cm}_mf^{jl}\hspace{0cm}_k+f^{jk}\hspace{0cm}_mf^{il}\hspace{0cm}_k=0,
\label{jacobian identity1}
\\
{\tilde f}_{ij}\hspace{0cm}^k{\tilde f}_{kl}\hspace{0cm}^m-{\tilde f}_{ik}\hspace{0cm}^m{\tilde f}_{jl}\hspace{0cm}^k+{\tilde f}_{jk}\hspace{0cm}^m{\tilde f}_{il}\hspace{0cm}^k=0,
\label{jacobian identity2}
\\
-f^{ij}\hspace{0cm}^k \tilde{f}_{lm}\hspace{0cm}^k+f^{ik}\hspace{0cm}_l \tilde{f}_{km}\hspace{0cm}^j-f^{jk}\hspace{0cm}_m \tilde{f}_{lk}\hspace{0cm}^i-f^{jk}\hspace{0cm}_l \tilde{f}_{km}\hspace{0cm}^i+f^{ik}\hspace{0cm}_m \tilde{f}_{lk}\hspace{0cm}^j=0,
\label{mix yacobi}
\end{eqnarray}
which are Jacobi identity of  ${ \cal G}$ and ${ \cal G}^*$ and mix Jacobi identity, respectively.
\par
{\bf Definition}:  \cite{GR1} A  3-Leibniz algebra ${ \cal A}$ equipped with a linear co commutator $\delta : { \cal A} \rightarrow
{ \cal A}\otimes{ \cal A} \otimes{ \cal A} $ defines a  \emph{3-Leibniz bialgebra} if:

{\it a} ) $\delta$ is a 1-cocycle of ${\cal A}$  and get  values on $ \otimes^3 {\cal A} $, i.e:
\begin{eqnarray}
\label{one cocycle}
\delta([T^{a},T^{b};T^{c}])&=&{ad^{(3)}}_{T^{b}\otimes T^{c}}\delta(T^{a}),
\end{eqnarray}
where  $\delta (T^{i})=\widetilde{f}^{lmn}\vspace{1mm}_{i}T_{l}\otimes T_{m}\otimes T_{n}$ and
\begin{eqnarray}
{ad^{(3)}}_{T^{b}\otimes T^{c}}&=&ad_{T^{b}\otimes T^{c}}\otimes 1\otimes 1+1\otimes ad_{T^{b}\otimes T^{c}}\otimes1+1\otimes1\otimes ad_{T^{b}\otimes T^{c}},
\end{eqnarray}
\\
with $\{T^a\}$s are  bases of 3-Leibniz algebra\footnote{ Note that  3-Leibniz algebra can be right or left 
 and according to left or right 3-Leibniz algebra  one can differently characterize the actions of ${\cal{A}}$ on $(\otimes)^3 \cal A$. We use only the right Leibniz algebra in this work, for details  see Ref.\cite{GR1}}  ${\cal A}$.

{\it b} ) 3-Leibniz bracket  can be defined as dual map $ ^t\delta:\otimes^3 {\cal A}^* \rightarrow  {\cal A}^*$ and it  is a commutator on $ \cal G^* $
\begin{equation}
(\tilde T_i\otimes \tilde T_k\otimes T_m,\delta(T^i))=(\delta^t(\tilde T_j\otimes \tilde T_k\otimes \tilde T_m),T^i)=\tilde f_{jkm}^i
\end{equation}
where $\{{\tilde T}_a\}$s are the bases for the space ${ \cal A}^*$ and  there is a natural pairing between ${\cal A}$ and $ {\cal A}^*$. Furthermore from this and (\ref{one cocycle}) we have the following relation between the structure constants \cite{GR1}:
\begin{equation}
\label{fund}
{f^{abc}}_{g}{\tilde{f}_{def}}\hspace{0cm}^{g}={f^{gbc}}_{f}{\tilde{f}_{deg}}\hspace{0cm}^{a}+{f^{gbc}}_{e}{\tilde{f}_{dfg}}\hspace{0cm}^{a}
-{f^{gbc}}_{d}{\tilde{f}_{efg}}\hspace{0cm}^{a}
\end{equation}
 \subsection{\label{An example}An example}
Generally, there is not Manin triple for 3-Leibniz bialgebras, but here we will show that there is Manin triple for a special example of 3-Leibniz bialgebras.

 {\bf Proposition: } {\em { $({\cal A}_{\cal G},{\cal A}^{*}_{\cal G^*})$ is a 3-Leibniz bialgebra if and only if  $(\cal G,\cal G^*)$ be a Lie  bialgebra}}\footnote{Note that in the forthcoming section, we will consider a Lie algebra $\cal G$ such that related Lie bialgebra $({\cal G},{\cal G}^*)$ has a 2-cocycle.}.
\par
Consider a special example of 3-Leibniz algebra ${\cal A}_{ \cal G}$ in  relation  with  Lie algebra $\cal G$.
This 3-Leibniz algebras ${\cal A}_{ \cal G}$ (mentioned in \cite{Mat} for a first time) have commutation relations as  follows\footnote{Note that for this 3-Leibniz algebra there is an antisymeticity only between elements $T^i$ and the  commutations such as $[T^{\pm},T^A,T^B]$ are not fully antisymetric.}:
\begin{eqnarray}
	\label{example 3-bracket}
	[T^{-},T^{a};T^{b}]=0,\hspace{1cm}
	[T^{+},T^{i};T^{j}]={f}^{ij}\hspace{0cm}_{k}T^{k},\hspace{1cm}
	[T^{i},T^{j};T^{k}]=-f^{ijk}T^{-},
\end{eqnarray}
where $\{T^i\}$s and  $f^{ij}\hspace{0cm}_k$ are basis  and  structure constant of the Lie algebra ${ \cal G}$  respectively with $[T^{i},T^{j}]={f}^{ij}\hspace{0cm}_{k}T^k$ and  fundamental identity for ${\cal A}_{ \cal G}$  has the following form: 
\begin{eqnarray}
f^{abc}\hspace{0cm}_ff^{fde}\hspace{0cm}_g- f^{ade}\hspace{0cm}_ff^{fbc}\hspace{0cm}_g-f^{bde}\hspace{0cm}_ff^{afc}\hspace{0cm}_g-f^{cde}\hspace{0cm}_ff^{abf}\hspace{0cm}_g=0,
\end{eqnarray}
where $a,b= i,,-,+$.

Now, we consider 3-Leibniz algebra structure on ${\cal A}^*_{\cal G^*}$:
\begin{eqnarray}
	\label{2example 3-bracket}
	[\tilde{T}_{-},\tilde{T}_{a};\tilde{T}_{b}]=0,\hspace{1cm}
	[\tilde{T}_{+},\tilde{T}_{i};\tilde{T}_{j}]={\tilde{f}}_{ij}\hspace{0cm}^{k}\tilde{T}_{k},\hspace{1cm}
	[\tilde{T}_{i},\tilde{T}_{j};\tilde{T}_{k}]=-\tilde{f}_{ijk}T_{-},
\end{eqnarray}
where  ${\tilde T_i}$ are bases for ${\cal G}^*$ with $[\tilde{T}_{i},\tilde{T}_{j}]={\tilde{f}}_{ij}\hspace{0cm}^{k}
\tilde{T}_{k}$, $i,j,k=1,2,...,dim { \cal G^*}$, where $ {\cal G}^*$  is the dual  Lie algebra of $\cal G$, and the fundamental identity for this 3-Leibniz algebra which is obtained as:
\begin{eqnarray}
{\tilde f}_{abc}\hspace{0cm}^f{\tilde f}_{fde}\hspace{0cm}^g- {\tilde f}_{ade}\hspace{0cm}^f{\tilde f}_{fbc}\hspace{0cm}^g-{\tilde f}_{bde}\hspace{0cm}^f{\tilde f}_{afc}\hspace{0cm}^g-{\tilde f}_{cde}\hspace{0cm}^f{\tilde f}_{abf}\hspace{0cm}^g=0,
\end{eqnarray} 
is equivalent to the Jacobi identity of the Lie algebra ${\cal G}^*$. By considering  the following inner product:
\begin{eqnarray}
\nonumber
(T^i,{\tilde T}_j)=\delta^i_j,\hspace{1cm}(T^i,T^j)=({\tilde T}_i,{\tilde T}_j)=0,
\end{eqnarray}
 one can show that the space ${\cal A}_{\cal G}\oplus {\cal A}^*_{{\cal G}^*}$  with the commutation relations (\ref{example 3-bracket},\ref{2example 3-bracket}) and continuing that, is a 3-Leibniz algebra if (${\cal A}_{\cal G},{\cal A}^*_{{\cal G}^*}$) is a 3-Leibniz bialgebra\footnote{Note that this is not direct sum of 3-Lie algebras.} and these are  equivalent if $({\cal G},{\cal G}^{*})$ construct a Lie bialgebra \cite{GR1}:
\begin{eqnarray}
\nonumber
[T^+,T^i;T^j]&=&f^{ij}\hspace{0cm}_kT^k,\hspace{.3cm}[T^i,T^j;T^k]=-f^{ijk}\hspace{0cm}_{-}T^-,\hspace{.3cm}[T^{\tilde +},T^{\tilde i};T^{\tilde j}]=\tilde{f}_{ij}\hspace{0cm}^kT^{\tilde k},\hspace{.3cm}[T^{\tilde i},T^{\tilde j};T^{\tilde k}]=\tilde{f}_{ijk}T^{\tilde -},
\\
\nonumber
[T^+,T^{\tilde i};T^j]&=&-f^{kj}\hspace{0cm}_iT^{\tilde k},\hspace{.3cm}[T^+,T^{\tilde i};T^{\tilde j}]=\tilde{f}_{kij}T^k,\hspace{.3cm}[T^i,T^{\tilde j};T^k]=-f^{ik}\hspace{0cm}_jT^{\tilde -},\hspace{.3cm}[T^i,T^{\tilde j};T^{\tilde k}]=-\tilde{f}_{jk}\hspace{0cm}^iT^-,
\\
\label{123456}
[T^{\tilde{+}},T^i;T^{\tilde j}]&=&-\tilde{f}_{kj}\hspace{0cm}^iT^k,\hspace{.3cm}[T^{\tilde +},T^i;T^j]=f^{kij},
\end{eqnarray}
 if we take  ${T^A}$ as a basis for the Manin triple of 3-Leibniz algebra with $A=i$, $T^i=T^i$ and $A= i+ {\textrm {dim} }{\cal G}+2$, $T^{ i+ {\textrm {dim} }{\cal G}+2}=\tilde{T}^i$  and $A= (-)+{\textrm {dim} }{\cal G}+2$, $T^{ (-)+{\textrm {dim} }{\cal G}+2}=T^{\tilde -}$ and $A=(+)+{\textrm {dim} }{\cal G}+2$, $T^{(+)+{\textrm {dim} }{\cal G}+2}=T^{\tilde +}$, these are not exactly commutation relations that showed in \cite{ali}. If we take $,a\equiv \tilde{+},d\equiv \tilde{+}$ and other indice from algebra then we will have:
\begin{eqnarray}
\label{1}
	f_{ijk}\tilde{f}^{lmk}+\tilde{f}^{kl}\hspace{0cm}_if_{kj}\hspace{0cm}^m+\tilde{f}^{km}\hspace{0cm}_if_{kj}\hspace{0cm}^l=0,
\end{eqnarray}
if $d\equiv -,c\equiv -$ 
\begin{equation}
\label{2}
f_{ki}\hspace{0cm}^m\tilde{f}^{lk}\hspace{0cm}_j=0,
\end{equation}
and we take $d\equiv \tilde{-}, c\equiv \tilde{-}$ will have
\begin{equation}
\label{3}
f_{ki}\hspace{0cm}^m\tilde{f}^{lk}\hspace{0cm}_j=0.
\end{equation}
Now, summation of (\ref{1}),(\ref{2}),(\ref{3}) gives us (\ref{mix yacobi})  which is mix Jacobi for Lie bialgebra.
 
\section{\label{BL model on Manin triple of 3-Lie algebra }BL model on Manin of 3-Leibniz algebras ( M2 $\leftrightarrow$ D2 ) }
In the previous section,  we  study   the Manin triple $({\cal D}, {\cal A}_{\cal G},{\cal A}_{{\cal G}^*})$ and correspondence between the 3-Leibniz bialgebra  $({\cal A}_{\cal G},{\cal A}_{{\cal G}^*})$ and Lie bialgebra $({\cal G},{\cal G}^*)$\footnote{ In general the Manin triple of 3- Leibniz bialgebra $({\cal D}, {\cal A}_{\cal G},{\cal A}_{{\cal G}^*})$ does  not exsit and Lie bialgebra $({\cal G},{\cal G}^*)$ have no correspondence with $({\cal A},{\cal A}^*)$, but for this  special example ${\cal D}$  is a 3-Leibniz algebra and  there is Manin triple $({\cal D}, {\cal A}_{\cal G},{\cal A}_{{\cal G}^*})$, then we have this correspondence.} in the special case. In this section, we will use the $N=6$  BL model which was explained in section 2 with one difference on the 3-Leibniz algebras. The 3-Leibniz algebras in our model is  ($\cal D$) , i.e.,  the Manin triple which  is $(4+2 {\textrm {dim} }{{\cal G}})$ dimensional  3-Leibniz algebra\footnote{ $Z^{\alpha}_a$ are the fields of model that are 3-Leibniz algebra ($\cal D$) valued and $(4 + 2 {\textrm{ dim }}G)$ is the dimension of  3-Leibniz
algebra $\cal D$.}. We will show its relation with WZW model with the bialgebraic structure equipped with one 2-cocycle. Note that, with this work all provided issues and conditions in the section \ref{sec:N=6 BLG}  for BL models are generalizable to a  Manin triple as a  3-Leibniz algebra, i.e., the shape of the equation of motions,  Lagrangian and ... are not modified. So consider the Lagrangian as follows:
\begin{eqnarray}
\nonumber
 {\cal L} &=& - D^\mu {\bar Z}_{\alpha}^A D_{\mu} Z^{\alpha}_A -
i{\bar\psi}^{\alpha A}\gamma^{\mu} D_{\mu}\psi_{\alpha A} -V+{\cal L}_{CS}
\\
\nonumber
&-& i F^{AB{\bar C}{\bar D}}{\bar\psi}^{\alpha}_{\bar D} \psi_{\alpha A}
Z^{\beta}_B{\bar Z}_{\beta{\bar C}}+2iF^{AB{\bar C}{\bar D}}{\bar\psi}^{\alpha}_{\bar D}\psi_{\beta A}Z^{\beta}_B{\bar Z}_{\alpha{\bar C}}
\\
\label{BLG action}
&+&\frac{i}{2}\varepsilon_{\alpha \beta \gamma \delta}F^{AB{\bar C}{\bar D}}{\bar\psi}^{\alpha}_{\bar D}\psi^{\beta}_{\bar C} Z^{\gamma}_AZ^{\delta}_B
-\frac{i}{2}\varepsilon^{\alpha \beta \gamma \delta}F^{CD{\bar A}{\bar B}}{\bar\psi}_{\alpha C}\psi_{\beta D}{\bar Z}_{\gamma {\bar A}}{\bar Z}_{\delta{\bar B}}\
\end{eqnarray}
 such that, structure constants have the following  relations  obtained from (\ref{123456}):
\begin{equation}
[T^A,T^B;T^C]=F^{ABC}\hspace{0cm}_DT^D,
\end{equation}
\begin{eqnarray}
\nonumber
&&F^{-AB}\hspace{0cm}_{C}=0,\hspace{1cm}F^{{\tilde-}AB}\hspace{0cm}_{C}=0,
\hspace{1cm}
F^{ABC}\hspace{0cm}_{+}=0,\hspace{1cm}F^{ABC}\hspace{0cm}_{{\tilde+}}=0,
\\
\nonumber
&&F^{+ij}\hspace{0cm}_k=f^{ij}\hspace{0cm}_k,\hspace{.5cm}F^{ijk}\hspace{0cm}_-=-f^{ijk},\hspace{.5cm}F^{\tilde{+}\tilde{i}\tilde{j}}\hspace{0cm}_{\tilde k}=\tilde{f}_{ij}\hspace{0cm}^k,\hspace{.5cm}F^{\tilde{i}\tilde{j}\tilde{k}}\hspace{0cm}_{\tilde -}=-\tilde{f}_{ijk},\hspace{.5cm}F^{+{\tilde i}j}\hspace{0cm}_{\tilde k}=-f^{kj}\hspace{0cm}_i,
\\
\label{1234567}
&&F^{+\tilde{i}\tilde{j}}\hspace{0cm}_k=\tilde{f}_{kij},\hspace{.5cm}F^{i\tilde{j}k}\hspace{0cm}_{\tilde -}=-f^{ik}\hspace{0cm}_j,\hspace{.5cm}F^{i\tilde{j}\tilde{k}}\hspace{0cm}_{-}=-\tilde{f}_{jk}\hspace{0cm}^i,\hspace{.5cm}F^{\tilde{+}i\tilde{j}}\hspace{0cm}_{k}=-\tilde{f}_{kj}\hspace{0cm}^i,\hspace{.5cm}F^{\tilde{+}ij}\hspace{0cm}_{\tilde k}=f^{kij}.
\end{eqnarray}
the gauge transformation for gauge fields  will be 
\begin{eqnarray}
\delta \tilde A_\mu{}^B{}_A=\partial_\mu \tilde\Lambda^B{}_A-
\tilde \Lambda^B{}_C \tilde A_\mu{}^A{}_A +\tilde A_\mu{}^B{}_C
\tilde \Lambda^C{}_A,\hspace{.3cm},
\end{eqnarray}
where ${\tilde \Lambda}_{\mu D}^C = {F^{CBA}}_D \Lambda_{\mu AB}$. By using the  eq.(34), we will have following relations:
\begin{eqnarray}
\label{11111}
\delta A_{\mu{\bar A}{\bar B}}=\partial_{\mu}\Lambda_2+[A_{\mu{+}{\bar D}},\Lambda_2]+[A_{\mu{\tilde +}{\bar D}},\Lambda_2]+[A_{\mu{\bar C}{\bar D}},\Lambda_1],
\\
\label{22222}
\delta A_{\mu{\tilde +}{\tilde i}}=\partial_{\mu}\Lambda_1+[A_{\mu{\tilde +}{\bar A}},\Lambda_1]+[A_{\mu{+}{\bar A}},\Lambda_1],
\\
\label{33333}
\delta A_{\mu{+}{\tilde i}}=\partial_{\mu}\Lambda_1+[A_{\mu{\tilde +}{\bar A}},\Lambda_1]+[A_{\mu{+}{\bar A}},\Lambda_1],
\end{eqnarray}
so that ${\Lambda}_1$ and ${\Lambda}_2$ take following relations:
\begin{equation}
 \label{gaugelst}
{\Lambda}_1= 2\Lambda_{+{\bar A}} T^{\bar A}\hspace{.2cm}\textrm{or}\hspace{.2cm} 2\Lambda_{{\tilde +}\bar A} T^{\bar A}, \qquad 
\Lambda_2 = \Lambda_{{\bar A}{\bar B}} F^{{\bar A}{\bar B}}\hspace{0cm}_{\bar C} T^{\bar C}.
\end{equation}
Note that we have used  $\bar A$ because we thought it could be confused with $A$ in 3-Leibniz bialgebra which could  get $i, \tilde i, +,\tilde+,-, \tilde- $ values, i.e.,  $\{T^{\bar A}\}$s are the basis of Manin triple of Lie bialgebra with ${\bar A}=i$, $T^i=T^i$, ${\bar A}= i+ {\textrm {dim} }{\cal G}$, $T^{ i+ {\textrm {dim} }{\cal G}}=\tilde{T}^i$ and $ F^{{\bar A}{\bar B}}\hspace{0cm}_{\bar C}$ is structure constant of Manin triple of Lie bialgebra (${\cal D}_{\cal G}$). The other index you will see in the following relation is $\cal A$. As you know when we separated the structure constants with only $i,\tilde i$, it must remain some of them with a combinational index of them and $+,\tilde +,-, \tilde -$ which  will be shown with $\cal A$ in the rest of this work.Also from eq.(\ref{gaugelst}) the gauge group for this case is ${\bf D}_{\cal G} \otimes {\bf D}_{\cal G}$ and after using the structure constants of 3-Leibniz bialgebra, it reduce to ${\bf D}_{\cal G}$, i.e., Drinfeld double of Lie bialgebra.  Now using of the above relations  the terms of Lagrangian turn into the following form: 
 \begin{eqnarray}
\nonumber
D_{\mu}Z^{\alpha}_DD^{\mu}{\bar Z}^D_{\alpha}&=&\partial_{\mu}Z^{\alpha}_{\bar D}\partial^{\mu}{\bar Z}^{\bar D}_{\alpha}+C_{\mu}^iC_i^{\mu}{\bar Z}^+_{\alpha}Z^{\alpha}_++4 C_{\mu}^if_{jki}A_{\mu}^k{\bar Z}^j_{\alpha}Z^{\alpha}_++4f^{ijk}f_{lmk}A_{\mu j}A^{\mu m}{\bar Z}^+_{\alpha}Z^{\alpha}_i+C_{\mu}^iC^{\mu}_j{\bar Z}^j_{\alpha}Z^{\alpha}_i
\\
&+&C_{\mu}^{'i}C_i^{'\mu}{\bar Z}^{\tilde +}_{\alpha}Z^{\alpha}_{\tilde +}+4 C_{\mu}^{'i}f_{jki}A_{\mu}^{'k}{\bar Z}^{\tilde j}_{\alpha}Z^{\alpha}_{\tilde +}+4f^{ijk}f_{lmk}A_{\mu j}^{'}A^{'\mu m}{\bar Z}^{\tilde +}_{\alpha}Z^{\alpha}_{\tilde i}+C_{\mu}^{'i}C^{'\mu}_j{\bar Z}^{\tilde j}_{\alpha}Z^{\alpha}_{\tilde i}+...
\end{eqnarray}
where $Z^{\alpha}=X^{\alpha}+iX^{\alpha +4}$. Furthermore, the first term of CS term  (\ref{CSC}) turns into the following forms:
 \begin{eqnarray}
\frac{1}{2}\epsilon^{\mu\nu\lambda}
F^{ABCD}A_{\mu AB}\partial_\nu A_{\lambda CD}=2\epsilon^{\mu\nu\lambda}
F^{{\bar B}{\bar C}{\bar D}}A_{\mu {\bar B}{\bar C}}\partial_\nu A_{\lambda {\bar D}}+2\epsilon^{\mu\nu\lambda}
F^{{\cal B}{\cal C}{\cal D}}A_{\mu {\cal B}{\cal C}}\partial_\nu A_{\lambda {\cal D}}
\label{CS11}
\end{eqnarray}
where 
 \begin{eqnarray}
\nonumber
&&\epsilon^{\mu\nu\lambda}F^{{\bar B}{\bar C}{\bar D}}A_{\mu {\bar B}}\partial_\nu A_{\lambda {\bar C}{\bar D}}=\frac{1}{3}\epsilon^{\mu\nu\lambda}f^{jk}\hspace{0cm}_{i}A_{\mu}^i\partial_{\nu}A_{\lambda jk}+\frac{2}{3}\epsilon^{\mu\nu\lambda}f^{ij}\hspace{0cm}_{k}A_{\mu i}\partial_{\nu}A_{\lambda j}^k+\frac{2}{3}\epsilon^{\mu\nu\lambda}f^{ki}\hspace{0cm}_{j}A_{\mu i}\partial_{\nu}A_{\lambda \tilde{j}}^{\tilde k}
\\
\nonumber
&+&\frac{2}{3}\epsilon^{\mu\nu\lambda}f^{ij}\hspace{0cm}_{k}A_{\mu}^{\tilde i}\partial_{\nu}A_{\lambda j\tilde{k}}+\frac{2}{3}\epsilon^{\mu\nu\lambda}f^{jk}\hspace{0cm}_{i}A_{\mu}^{\tilde i}\partial_{\nu}A_{\lambda j}^{\tilde k}+\frac{1}{3}\epsilon^{\mu\nu\lambda}{\tilde f}_{jk}\hspace{0cm}^{i}A_{\mu}^{\tilde i}\partial_{\nu}A_{\lambda {\tilde j}{\tilde k}}+\frac{2}{3}\epsilon^{\mu\nu\lambda}{\tilde f}_{ik}\hspace{0cm}^{j}A_{\mu}^{i}\partial_{\nu}A_{\lambda {j}{\tilde k}}
\\
&+&\frac{2}{3}\epsilon^{\mu\nu\lambda}{\tilde f}_{jk}\hspace{0cm}^{i}A_{\mu}^{i}\partial_{\nu}A_{\lambda{\tilde k}}^j+\frac{2}{3}\epsilon^{\mu\nu\lambda}{\tilde f}_{k{\tilde i}}\hspace{0cm}^{j}A_{\mu\tilde i}\partial_{\nu}A_{\lambda {j}}^k+\frac{2}{3}\epsilon^{\mu\nu\lambda}{\tilde f}_{ij}\hspace{0cm}^{k}A_{\mu\tilde i}\partial_{\nu}A_{\lambda {\tilde j}}^{\tilde k}
\label{WZW1}
\end{eqnarray}
and 
 \begin{eqnarray}
\nonumber
\epsilon^{\mu\nu\lambda}
F^{{\cal B}{\cal C}{\cal D}}A_{\mu {\cal B}}\partial_\nu A_{\lambda {\cal C}{\cal D}}&=&-\frac{1}{2}\epsilon^{\mu\nu\lambda}f^{kj}\hspace{0cm}_iA_{\mu j{\tilde i}}\partial_{\nu}A_{\lambda}^{
	\tilde k}-\frac{1}{3}\epsilon^{\mu\nu\lambda}f^{jk}\hspace{0cm}_i A_{\mu {\tilde i}}^{\tilde j}\partial_{\nu}A_{\lambda k}-\epsilon^{\mu\nu\lambda}{\tilde f}_{jik}A_{\mu \tilde i}^j\partial_{\nu}A_{\lambda \tilde k}
\\
\nonumber
&+&\frac{5}{6}\epsilon^{\mu\nu\lambda}{\tilde f}_{ijk}A_{\mu {\tilde j}{\tilde i}}\partial_{\nu}A_{\lambda}^k-\frac{5}{6}\epsilon^{\mu\nu\lambda}f^{ijk}A_{\mu ji}\partial_{\nu}A_{\lambda}^{\tilde k}-\frac{5}{6}\epsilon^{\mu\nu\lambda}f^{ikj}A_{\mu i}^{\tilde j}\partial_{\nu}A_{\lambda k}
\\
\nonumber
&+&\frac{1}{3}\epsilon^{\mu\nu\lambda}{\tilde f}_{jk}\hspace{0cm}^iA_{\mu i}^j\partial_{\nu}A_{\lambda \tilde k}-\frac{5}{6}\epsilon^{\mu\nu\lambda}{\tilde f}_{ij}\hspace{0cm}^kA_{\mu {\tilde j}{\tilde i}}\partial_{\nu}A_{\lambda}^{\tilde k}+\frac{5}{6}\epsilon^{\mu\nu\lambda}{\tilde f}_{ji}\hspace{0cm}^kA_{\mu}^j\partial_{\nu}A_{\lambda {\tilde i}k}
\\
&-&\frac{1}{2}\epsilon^{\mu\nu\lambda}{\tilde f}_{ki}\hspace{0cm}^jA_{\mu j}\partial_{\nu}A_{\lambda \tilde i}^k+\frac{2}{3}\epsilon^{\mu\nu\lambda}f^{jk}\hspace{0cm}_iA_{\mu \tilde i}\partial_{\nu}A_{\lambda j}^{\tilde k}
 \end{eqnarray}
and the second term of CS term turn into 
 \begin{eqnarray}
\frac{1}{3}\epsilon^{\mu\nu\lambda}\hspace{-.08cm}F^{AEF}\hspace{-.08cm}_{G}\,F^{BCDG}\,A_{\mu AB}A_{\nu CD}A_{\lambda EF}\hspace{-.1cm}
=\hspace{-.1cm}-2\epsilon^{\mu\nu\lambda}\,F^{{\bar A}{\bar B}{\bar C}} F^{{\bar E}{\bar F}}\hspace{-.08cm}_{\bar A} A_{\mu {\bar E}{\bar F}}A_{\nu {\bar B}}A_{\lambda {\bar C}}-2\epsilon^{\mu\nu\lambda}\,F^{{\cal A}{\cal B}{\cal C}} F^{{\cal E}{\cal F}}\hspace{-.08cm}_{\cal A} A_{\mu {\cal E}{\cal F}}A_{\nu {\cal B}}A_{\lambda {\cal C}}
\label{CS22}
 \end{eqnarray}
where
 \begin{eqnarray}
 \nonumber
& &\hspace{-1cm}\epsilon^{\mu\nu\lambda}\hspace{-.09cm}F^{{\bar A}{\bar B}{\bar C}}\hspace{-.08cm} F^{{\bar E}{\bar F}}\hspace{-.08cm}_{\bar A} A_{\mu {\bar E}{\bar F}}A_{\nu {\bar B}}A_{\lambda {\bar C}}=\frac{1}{2}\epsilon^{\mu\nu\lambda}f^{jk}\hspace{0cm}_if^{il}\hspace{0cm}_m A_{\mu jk} A_{\nu l}A_{\lambda}^m+\frac{1}{2}\epsilon^{\mu\nu\lambda}f^{ij}\hspace{0cm}_kf^{lm}\hspace{0cm}_iA_{\mu j}^kA_{\nu l}A_{\lambda m}
\\
 \nonumber
&&\hspace{-1cm}+\epsilon^{\mu\nu\lambda}f^{ij}\hspace{0cm}_k\tilde{f}_{mi}\hspace{0cm}^lA_{\mu j{\tilde k}}A_{\nu l}A_{\lambda}^m+\frac{1}{2}\epsilon^{\mu\nu\lambda}f^{jk}\hspace{0cm}_i\tilde{f}_{lm}\hspace{0cm}^iA_{\mu jk}A_{\nu {\tilde l}}A_{\lambda}^m+\epsilon^{\mu\nu\lambda}f^{ij}\hspace{0cm}_k\tilde{f}_{il}\hspace{0cm}^mA_{\mu j}^kA_{\nu {\tilde l}}A_{\lambda m}+\epsilon^{\mu\nu\lambda}f^{ij}\hspace{0cm}_k f^{lm}\hspace{0cm}_iA_{\mu j{\tilde k}}A_{\nu}^{\tilde l}A_{\lambda m}
\\
 \nonumber
 &&\hspace{-1cm}+\frac{1}{2}\epsilon^{\mu\nu\lambda}f^{jk}\hspace{0cm}_i f^{mi}\hspace{0cm}_lA_{\mu j}^{\tilde k}A_{\nu {\tilde l}}A_{\lambda m}+\epsilon^{\mu\nu\lambda}f^{mi}\hspace{0cm}_l\tilde{f}_{ki}\hspace{0cm}^jA_{\mu j{\tilde k}}A_{\nu}^lA_{\lambda m}+\epsilon^{\mu\nu\lambda}f^{lm}\hspace{0cm}_i\tilde{f}_{kj}\hspace{0cm}^iA_{\mu{\tilde k}}^jA_{\nu l}A_{\lambda m}+\frac{1}{2}\epsilon^{\mu\nu\lambda}{\tilde f}_{jk}\hspace{0cm}^i\tilde{f}_{mi}\hspace{0cm}^lA_{\mu {\tilde j}{\tilde k}}A_{\nu l}A_{\lambda}^m
 \\
 \nonumber
&&\hspace{-1cm}+\epsilon^{\mu\nu\lambda}{\tilde f}_{ki}\hspace{0cm}^j\tilde{f}_{lm}\hspace{0cm}^iA_{\mu {\tilde j}{\tilde k}}A_{\nu{\tilde l}}A_{\lambda}^m+\epsilon^{\mu\nu\lambda}{\tilde f}_{kj}\hspace{0cm}^i\tilde{f}_{il}\hspace{0cm}^mA_{\mu{\tilde k}}^jA_{\nu{\tilde l}}A_{\lambda m}+\frac{1}{2}\epsilon^{\mu\nu\lambda}f^{im}\hspace{0cm}_l\tilde{f}_{ij}\hspace{0cm}^kA_{\mu{\tilde j}}^{\tilde k}A_{\nu {\tilde l}}A_{\lambda m}+\frac{1}{4}\epsilon^{\mu\nu\lambda}f^{lm}\hspace{0cm}_i\tilde{f}_{jk}\hspace{0cm}^iA_{\mu {\tilde j}{\tilde k}}A_{\nu l}A_{\lambda}^{\tilde m}
 \\
\nonumber
&&\hspace{-1cm}+\epsilon^{\mu\nu\lambda}f^{mi}\hspace{0cm}_l\tilde{f}_{ki}\hspace{0cm}^jA_{\mu j{\tilde k}}A_{\nu{\tilde l}}A_{\lambda}^{\tilde m}+\frac{1}{2}\epsilon^{\mu\nu\lambda}{\tilde f}_{jk}\hspace{0cm}^i\tilde{f}_{mi}\hspace{0cm}^lA_{\mu {\tilde j}{\tilde k}}A_{\nu}^{\tilde l}A_{\lambda {\tilde m}}+\frac{1}{2}\epsilon^{\mu\nu\lambda}{\tilde f}_{ij}\hspace{0cm}^k\tilde{f}_{lm}\hspace{0cm}^iA_{\mu {\tilde j}}^{\tilde k}A_{\nu{\tilde l}}A_{\lambda {\tilde m}}+\frac{1}{2}\epsilon^{\mu\nu\lambda}f^{jk}\hspace{0cm}_i\tilde{f}_{li}\hspace{0cm}^mA_{\mu j}^{\tilde k}A_{\nu l}A_{\lambda m}
\\
&+&\frac{1}{2}\epsilon^{\mu\nu\lambda}f^{jk}\hspace{0cm}_i f^{mi}\hspace{0cm}_lA_{\mu jk}A_{\nu{\tilde l}}A_{\lambda}^{\tilde m}+\epsilon^{\mu\nu\lambda}f^{ij}\hspace{0cm}_k\tilde{f}_{il}\hspace{0cm}^mA_{\mu j{\tilde k}}A_{\nu{\tilde l}}A_{\lambda}^{\tilde m}+\frac{1}{2}\epsilon^{\mu\nu\lambda}f^{jk}\hspace{0cm}_i\tilde{f}_{lm}\hspace{0cm}^iA_{\mu j{\tilde k}}A_{\nu{\tilde l}}A_{\lambda {\tilde m}},
 \label{WZW2}
   \end{eqnarray}
and
 \begin{eqnarray}
\nonumber
& &\hspace{-1.2cm}\epsilon^{\mu\nu\lambda}\,F^{{\cal A}{\cal B}{\cal C}} F^{{\cal E}{\cal F}}\hspace{-.08cm}_{\cal A} A_{\mu {\cal E}{\cal F}}A_{\nu {\cal B}}A_{\lambda {\cal C}}=\frac{1}{4}\epsilon^{\mu\nu\lambda}[(2{\tilde f}_{ijk}f^{pk}\hspace{0cm}_l-3f^{pk}\hspace{0cm}_i{\tilde f}_{pjl}-{\tilde f}_{ji}\hspace{0cm}^p{\tilde f}_{pl}\hspace{0cm}^k)A_{\nu{\tilde j}{\tilde i}}A_{\mu k}A_{\lambda}^l
\\
\nonumber
&&\hspace{-1.2cm}+(3{\tilde f}_{ij}\hspace{0cm}^pf^{kl}\hspace{0cm}_p-3{\tilde f}_{ijp}f^{plk})A_{\mu}^{\tilde k}A_{\nu {\tilde j}{\tilde i}}A_{\lambda l}
+
(3{\tilde f}_{ijp}{\tilde f}_{kl}\hspace{0cm}^p-3{\tilde f}_{il}\hspace{0cm}^p{\tilde f}_{pjk}-2{\tilde f}_{ij}\hspace{0cm}^p{\tilde f}_{plk}+{\tilde f}_{ilp}f_{kj}\hspace{0cm}^p)A_{\mu}^kA_{\nu {\tilde j}{\tilde i}}A_{\lambda l}
\\
\nonumber
&&\hspace{-1.2cm}+(-2{\tilde f}_{klp}f^{jp}\hspace{0cm}_i-{\tilde f}_{kip}f^{jp}\hspace{0cm}_l-{\tilde f}_{ilp}f^{pj}\hspace{0cm}_k)A_{\mu {\tilde k}}{\tilde i}A_{\nu}^{\tilde j}A_{\tilde l}
+
(f^{pl}\hspace{0cm}_if^{kj}\hspace{0cm}_p-f^{jlp}{\tilde f}_{ip}^k)A_{\mu \tilde i}^{\tilde k}A_{\nu j}A_{\lambda l}
\\
\nonumber
&&\hspace{-1.2cm}+(-{\tilde f}_{ip}\hspace{0cm}^k f^{pk}\hspace{0cm}_j-f^{kp}\hspace{0cm}_i{\tilde f}_{pj}\hspace{0cm}^l-f^{lp}\hspace{0cm}_i{\tilde f}_{pj}\hspace{0cm}^k-{\tilde f}_{ijp}f^{plk})A_{\mu {\tilde i}}A_{\nu \tilde j}A_{\lambda l}-f^{jk}\hspace{0cm}_i{\tilde f}_{lm}\hspace{0cm}^iA_{\mu jk}A_{\nu \tilde l}A_{\lambda}^m
\\
\nonumber
&&\hspace{-1.2cm}+({\tilde f}_{ijp}f^{lp}\hspace{0cm}_k-{\tilde f}_{kjp}f^{pl}\hspace{0cm}_i -{\tilde f}_{pk}\hspace{0cm}^l{\tilde f}_{ij}\hspace{0cm}^p)A_{\mu \tilde i}A_{\nu \tilde j}A_{\lambda \tilde k}^{\tilde l}
+
(f^{kj}\hspace{0cm}_pf^{pi}\hspace{0cm}_l+f^{ikp}{\tilde f}_{lp}\hspace{0cm}^j+2f^{ijp}{\tilde f}_{lp}\hspace{0cm}^k)A_{\mu k}A_{\nu ji}A_{\lambda}^l
\\
\nonumber
&&\hspace{-1.2cm}+(-3f^{ij}\hspace{0cm}_pf^{plk}+2f^{il}\hspace{0cm}_pf^{jpk}+f^{pl}\hspace{0cm}_if^{kj}\hspace{0cm}_p+2f^{ijp}f^{kl}\hspace{0cm}_p+f^{ilp}f^{kj}\hspace{0cm}_p)A_{\mu}^{\tilde k}A_{\nu ji}A_{\lambda l}
\\
\nonumber
&&\hspace{-1.2cm}+(2f^{ij}\hspace{0cm}_p{\tilde f}_{kl}\hspace{0cm}^p+{\tilde f}_{pl}\hspace{0cm}^if^{pj}\hspace{0cm}_k+2f^{ijp}{\tilde f}_{pkl})A_{\mu \tilde k}A_{\nu ji}A_{\lambda}^l+f^{pj}\hspace{0cm}_i{\tilde f}_{pl}\hspace{0cm}^kA_{\mu}A_{\nu j\tilde i}A_{\nu \tilde l}-2{\tilde f}_{ki}\hspace{0cm}^jf^{mi}\hspace{0cm}_lA_{\mu j}^kA_{\nu \tilde l}A_{\lambda m}
\\
\nonumber
&&\hspace{-1.2cm}+(-3{\tilde f}_{pj}\hspace{0cm}^if^{lkp}-2f^{ikp}{\tilde f}_{pj}\hspace{0cm}^l)A_{\mu ki}A_{\nu \tilde j}A_{\lambda}^{\tilde l}-2{\tilde f}_{ki}\hspace{0cm}^j f^{mi}_l A_{\mu j\tilde k}A_{\nu \tilde l}A_{\lambda}^{\tilde m}-f^{ki}\hspace{0cm}_jf^{lm}\hspace{0cm}_iA_{\mu \tilde j}^kA_{\nu l}A_{\lambda m}
\\
\nonumber
&&\hspace{-1.2cm}+(f^{li}\hspace{0cm}_pf^{pj}\hspace{0cm}_k+f^{ljp}{\tilde f}_{kp}\hspace{0cm}^i)A_{\mu i}^kA_{\nu j}A_{\lambda l}
+
({\tilde f}_{pk}\hspace{0cm}^lf^{ip}\hspace{0cm}_j+f^{lp}_j{\tilde f}_{pk}^i+f^{ilp}{\tilde f}_{pjk})A_{\mu i}^k A_{\nu \tilde j}A_{\lambda l}
\\
\nonumber
&&\hspace{-1.2cm}+(-f^{pl}\hspace{0cm}_k{\tilde f}_{jp}\hspace{0cm}^i-f^{il}\hspace{0cm}_p{\tilde f}_{jk}\hspace{0cm}^p+2{\tilde f}_{pk}\hspace{0cm}^if^{pl}\hspace{0cm}_j+2{\tilde f}_{pj}\hspace{0cm}^lf^{pi}\hspace{0cm}_k)A_{\mu {\tilde k}i}A_{\nu}^j A_{\lambda l}+(-{\tilde f}_{ij}\hspace{0cm}^p{\tilde f}_{lkp}+{\tilde f}_{kip}{\tilde f}_{lj}\hspace{0cm}^p-{\tilde f}_{ki}\hspace{0cm}^p {\tilde f}_{pjl})A_{\mu \tilde i}A_{\nu \tilde j}A_{\lambda \tilde k}^l
\\
\nonumber
&&\hspace{-1.2cm}+(-f^{ij}\hspace{0cm}_pf^{lp}\hspace{0cm}_k+f^{kj}\hspace{0cm}_pf^{lp}\hspace{0cm}_i+2f^{ijp}{\tilde f}_{kp}\hspace{0cm}^l-{\tilde f}_{kjp}f^{ipl}-f^{plj}{\tilde f}_{pk}\hspace{0cm}^i+2{\tilde f}_{pk}\hspace{0cm}^if^{jpl})A_{\mu {\tilde k}i}A_{\nu j}A_{\lambda}^{\tilde l}
\\
\nonumber
&&\hspace{-1.2cm}+(-{\tilde f}_{kp}\hspace{0cm}^i{\tilde f}_{jl}\hspace{0cm}^p-3f^{pi}\hspace{0cm}_k{\tilde f}_{pjl}-3{\tilde f}_{kjp}f^{ip}\hspace{0cm}_l+f^{ip}\hspace{0cm}_j{\tilde f}_{kpl})A_{\mu {\tilde k}i}A_{\nu \tilde j}A_{\lambda}^l+(-f^{ijp}{\tilde f}_{plk}+2{\tilde f}_{pl}\hspace{0cm}^jf^{ipk}-{\tilde f}_{pl}\hspace{0cm}^if^{pjk}\\
\nonumber
&&\hspace{-1.2cm}-2f^{ik}\hspace{0cm}_pf^{pj}\hspace{0cm}_l)A_{\mu i}^{\tilde k}A_{\nu j}A_{\lambda \tilde l}
+({\tilde f}_{pj}\hspace{0cm}^if^{lp}\hspace{0cm}_k+{\tilde f}_{kj}\hspace{0cm}^pf^{li}\hspace{0cm}_p+{\tilde f}_{kjp}f^{pil}-{\tilde f}_{kjp}f^{ipl}+3f^{pi}\hspace{0cm}_k{\tilde f}_{pj}\hspace{0cm}^l)A_{\mu {\tilde k}i}A_{\nu \tilde j}A_{\lambda}^{\tilde l}
\\
\nonumber
&&\hspace{-1.2cm}+(-f^{ij}\hspace{0cm}_pf^{plk}+f^{ijp}f^{kl}\hspace{0cm}_p-f^{jl}\hspace{0cm}_pf^{ipk})A_i^{\tilde k}A_{\nu j}A_{\lambda l}+f^{ji}\hspace{0cm}_k{\tilde f}_{li}\hspace{0cm}^mA_{\mu \tilde k}^{\tilde j}A_{\nu \tilde l}A_{\lambda m}-2{\tilde f}_{kj}\hspace{0cm}^i{\tilde f}_{mi}\hspace{0cm}^lA_{\mu j}^{\tilde k}A_{\nu l}A_{\lambda \tilde m}
\\
\nonumber
&&\hspace{-1.2cm}+({\tilde f}_{pj}\hspace{0cm}^if^{pk}\hspace{0cm}_l+{\tilde f}^{lj}\hspace{0cm}_pf^{ipk}-{\tilde f}_{jlp}f^{ipk}+f^{ki}\hspace{0cm}_p{\tilde f}_{jl}\hspace{0cm}^p)A_{\mu i}^{\tilde k}A_{\nu \tilde j}A_{\lambda \tilde l}
+(f^{pj}\hspace{0cm}_i{\tilde f}_{plk}-f^{pj}\hspace{0cm}_l{\tilde f}_{ipk}-{\tilde f}_{ilp}f^{pj}\hspace{0cm}_k+{\tilde f}_{ik}\hspace{0cm}^p{\tilde f}_{lp}\hspace{0cm}^j)A_{\mu {\tilde i}}^kA_{\nu j}A_{\lambda \tilde l}].
\\
 \end{eqnarray}
 Now,  if we assume that $Z^{\alpha}_+=Z^{\alpha}_{\tilde +}$ and also $A_{\mu +{\bar B}}=A_{\mu {\tilde +}{\bar B}}=A_{\mu {\bar B}}$\footnote{ In our model the gauge group is ${\bf D}_{\cal G}\otimes {\bf D}_{\cal G}$ which partial break to ${\bf D}_{\cal G}$. Forthermore we apply the assumption $A_{+\mu{\bar B}}=A_{{\tilde +}\mu{\bar B}}=A_{\mu{\bar B}}$ for obtaining and simplifying  the result to (\ref{FB}) if one not apply such assumption then in (\ref{FB}) further terms related to gauge group fields terms will appeard. Mean while according to (\ref{1234567}-\ref{gaugelst}) this assumption reduce the number of gauge field but not  the gauge group ${\bf D}_{\cal G}$.}, $F^{{\bar A}{\bar B}}\hspace{0cm}_{\bar C} A_{\mu {\bar A}{\bar B}}\equiv C_{\mu {\bar C}}$ and $F^{{\cal A}{\cal B}}\hspace{0cm}_{\cal C} A_{\mu {\cal A}{\cal B}}\equiv {\tilde C}_{\mu {\cal C}}$, we will have the following relation from CS term on the special 3-Leibniz algebra $\cal D$:
\begin{eqnarray}
\nonumber
{\cal L}_{CS}=\frac{1}{2}\epsilon^{\mu\nu\lambda}\{C_{\mu {\bar B}}(\partial_{\nu}A_{\lambda }^{\bar B}-\partial_{\lambda}A_{\nu}^{\bar B}-[A_{\nu },A_{\lambda }]_{\bar B} )&+&{\tilde C}_{\mu {\cal B}}(\partial_{\nu}A_{\lambda }^{\cal B}-\partial_{\lambda}A_{\nu}^{\cal B}-[A_{\nu },A_{\lambda }]_{\cal B} ).
\end{eqnarray}
Finally, the  $N=6$ BL Lagrangian (\ref{BLG action}) is rewritten as follows:
\begin{eqnarray}
\nonumber
{\cal L} =  \frac{1}{2}D_{\mu}X^{{\bar A}(I)}D^{\mu}X_{\bar A}^{(I)}-2 g_{YM}^2 C_{\mu}^{\bar B}C^{\mu}_{\bar B}-2 g_{YM}^2 {\tilde C}_{\mu}^{\cal B}{\tilde C}^{\mu}_{\cal B}
+ 2\,\epsilon^{\mu\nu\lambda}\,
  C_{\mu {\bar A}} B_{\nu\lambda}^{\bar A} +  2\,\epsilon^{\mu\nu\lambda}\,
    {\tilde C}_{\mu}^{\cal B} F_{\nu\lambda {\cal B}}+...,
\label{L2}
\end{eqnarray}
where $I=1,...,8$ and the ... terms do not affect the Yang-Mills and WZW-like terms.
 In this way, one can  reach  to Yang-Mills relation (similar to \cite{Matsu}) with 7 scalar fields by giving vacuum expectation value (VEV)  to  real part of one of complex scalar fields $Z^{\alpha}$, in this manner  imaginary part of it  behaves as a Goldestone mode in Higges mechanism \cite{EMP}. 
Now, by defining $B_{\nu\lambda {\bar A}}=\partial_{\nu}A_{\lambda {\bar A}}-\partial_{\lambda}A_{\nu {\bar A}}-[A_{\nu},A_{\lambda}]_{\bar A}
$, $F_{\nu\lambda }^{\cal A}=\partial_{\nu}A_{\lambda}^{\cal A}-\partial_{\lambda}A_{\nu}^{\cal A}-[A_{\nu},A_{\lambda}]^{\cal A}$; by  integrating  $C_{\mu B}$, $ \tilde{C}_{\mu}^{\cal B}$   in (\ref{L2})  we will obtain the following equation:
\begin{equation}
\label{FB}
{\cal L}=\frac{1}{4}F_{\nu\lambda {\cal A}}F^{\nu\lambda {\cal A}}+\frac{1}{8}B_{\nu\lambda \bar A}B^{\nu\lambda \bar A}+...,
\end{equation}
such that $F_{\nu\lambda {\cal A}}$ and $B_{\nu\lambda \bar A}$ are two strength fields on ${\cal D}_{\cal G} \oplus{\cal D}^*_{{\cal G}^*}$ which look like Yang-Mills and B-field of a string, respectively. Note that $B_{\mu \lambda \bar A}$ as the antisymmetric Lie algebra ${\cal D}_{\cal G}$ valued field obtained from Higgs mechanism while CS term is written to the BL model on Manin triple $\cal D$ not from prior 3-form since the BL Lagrangian do not conclude it. A 3-form can be written as $H=dB+C$ for the M2-M5 system as s mentioned in \cite{chu}, although there is not a 3-form $C$ in this case, a 2-form $B$ on the boundary of M2-brane ending to M5-brane could be the one appeared in our model. The Lagrangian (\ref{FB}) is a conclusion from low energy limit of DBI action related to the propagating of the string in the background $G_{\mu\nu}$ and Lie algebra valued fields $A_{\mu\nu A}$ and $B_{\mu \nu A}$. So the antisymmetric Lie algebra valued form $B_{\mu\nu A}$ could be B-field of the string at the boundary of the M2-brane end to M5-brane \cite{chu}. This means we have considered boundary conditions of the M2-M5 system which make some restrictions on the gauge fields. Going to the boundary for the supersymmetric problems change the number of supersymmetry so that for the $N=6$ supersymmetry two anti-chiral fermions on the boundary is disappeared. So we could expect $N=(4,4)$ supersymmetry to the string \cite{Howe}\footnote{Note that, this method for obtaining of WZW is different of the method mentioned in \cite{Smith,Okazaki,Moore}.}. In this way, the $N=6$ BL model related to the bosonic sector of a $N=(4,4)$ supersymmetric string theory in our model. The $N=(4,4)$ supersymmetric WZW model is constructed from the group of Manin triple (as a Lie algebra) of Lie bialgebra, i.e., Drinfeld double ${\bf D}_{\cal G}$ with one 2-cocycle. Applying 3-Leibniz bialgebra in the BL model make clear relation between two superconformal theory (superconformal BLG model stated in Ref.\cite{Farrill,Showrz}) and one can obtain them from each other, i.e., BL model with 3-Leibniz algebra structure $\cal D$ as a Manin triple is constructed by having 2-dimensional $N=(4,4)$ supersymmetric WZW models (analyzed \cite{AR}) with Lie bialgebra structure equipped one 2-cocycle and vice versa. We meant the vice versa process is only from the algebraic point of view. On the other hand, the algebraic relation between Manin triple of 3- Leibniz algebra and Lie bialgebra (as we describe in the proposition) make possible constructing BL model from the $N=(4,4)$ supersymmetric WZW model. There exist two gluing matrices $R,\bar R$ in an $N=(2,2)$ WZW which must satisfy $R \in Aut (\cal G)$ and ${\bar R}\in Hom({\cal G}_-,{\cal G}_+)$\footnote{Note that ${\cal G}_+$ and ${\cal G}_-$ are Lie algebra and dual of it, respectivly, that make the Lie bialgebra ${\cal G}$, i.e. ${\cal G}={\cal G}_{+}\oplus {\cal G}_{-}$.}, then according to these gluing matrices there are two types D-brane, i.e., B-type and A-type respectively. In order to preserve half of the bulk supersymmetry \cite{Zabzin} then it is natural to conclude that for the $N=(4,4)$ WZW. In this case, existence one 2-cocycle which can play an important role, also possible relations between gluing matrices $R$ and $\bar R$ according to \cite{Parkhomenko} and \cite{AR}.

Now, it remains to show that $B_{\mu \nu k}$ is a B-field. As we said a self-dual string appears on the boundary of M5-brane which will construct a 3-form on the worldvolume. Appeared B-field in our model is a result of 3-form that play background field role. Born-Infeld action for propagated string in a background with $G_{\mu\nu},A_{\mu}$ and $ B_{\mu\nu}$ fields resulting action is DBI action \cite{Liegh,Tong}:
\begin{eqnarray}
\nonumber
S&=&\int d^{p+1}\zeta \sqrt{-det(G_{\mu\nu}h_{AB}+2\pi\alpha^{\prime}F_{\mu\nu AB}+B_{\mu\nu AB})},
     \\
&=&\frac{1}{2}F_{\nu\lambda K}F^{\nu\lambda K}+\frac{1}{4}B_{\nu\lambda K}B^{\nu\lambda K}+... .
\label{mas}
 \end{eqnarray}
We must show that this B-field can be related to the B-field of  WZW action ( as in Ref.\cite{ali}), for this reason, we  use Lie algebra valued fields:
\begin{eqnarray}
S_{WZW-like}=\int d^{3}x\epsilon^{\alpha\beta\gamma}L_{\mu}\hspace{0cm}^{L}L_{\nu}\hspace{0cm}^{M}L_{\lambda}^{N}\partial_{\alpha}X^{I\mu}\partial_{\beta}X^{J\nu}\partial_{\gamma}X^{K\lambda} Tr([T_{I}T_{L},T_{J}T_{M}],T_{K}T_{N}),
 \end{eqnarray}
by some calculations we obtain
\begin{eqnarray}
\nonumber
S_{WZW-like}&=&\int
d^{2}x\ \{\frac{1}{6}\epsilon^{\beta\gamma}{B_{\nu\mu}}^{Q}\partial_{\beta}X^{J\nu}\partial_{\gamma}X^{I\mu}Tr(T_{J}T_{I}T_{Q})
+\frac{1}{6}\epsilon^{\alpha\gamma}{B_{\nu\lambda}}^{Q}\partial_{\alpha}X^{J\nu}\partial_{\gamma}X^{K\lambda}Tr(T_{J}T_{K}T_{Q})
\\
&+&\frac{1}{6}\epsilon^{\alpha\gamma}{B_{\mu\lambda}}^{Q}\partial_{\alpha}X^{I\mu}\partial_{\gamma}X^{K\lambda}Tr(T_{I}T_{K}T_{Q})\}+... 
,
\label{WZW  action noncordinate basis}
\end{eqnarray}
where  ${B_{\nu\mu}}^{Q}={L_{\nu}}\hspace{0cm}^{L}{L_{\lambda}}\hspace{0cm}^{N}{f_{NL}}^{P}x^J {f_{PJ}}^{Q}$, $L_{\mu}^LX^{I\mu}T_IT_L|_{boundry}=x^LT_L|_{boundry}$.

\section{Boundary conditions of BL model on Manin triple of 3-Leibniz algebra }
In this section, we obtain the boundary conditions of $N=6$ BL model with 3-Leibniz bialgebra structure, as we have shown for the $N=8$ BLG model. The difference between them is in the number of supersymmetry and structure constants which are not totally antisymmetric. For this purpose, we must be careful in decreasing the number of supersymmetries. The normal component of supercurrent to boundary direction must be discarded in order to preserve maximum unbroken supersymmetry, i.e., we obtain the supercurrent and reset to zero normal component of it. For this purpose, we will need the other presentation of BL model which have been applied by Bagger and Lambert in Ref.\cite{Bag4}. They have shown that $F^{AB{ C}{ D}}$ sets up a Lie algebra $\mathcal G$ with the following form:
\begin{equation}
\mathcal{G} = \otimes_\lambda \mathcal{G}_\lambda
\end{equation}
such that $\mathcal G_{\lambda}$s are commuting Lie algebras and this allows one to rewrite  the   Lagrangian (\ref{BLG action}) as follows: 
\begin{eqnarray}
\nonumber
L &= &-Tr (D^\mu {\bar Z}_{\alpha} , D_\mu Z^{\alpha}) - i Tr ({\bar {\psi}}^{\alpha} , \gamma^\mu D_\mu \psi_{\alpha}) - V + L_{CS} \\\nonumber
&-& i Tr ({\bar {\psi}}^{\alpha} , [\psi_{\alpha} , Z^{\beta} ; {\bar Z}_{\beta}]) + 2i Tr ({\bar \psi}^{\alpha}, [\psi_{\beta} , Z^{\beta} ; {\bar Z}_{\alpha} ])\\ 
&+ &\frac{i}{2} \varepsilon_{\alpha \beta \gamma \delta} Tr ({\bar {\psi}}^{\alpha} , [Z^{\gamma} , Z^{\delta} ; \psi^{\alpha} ]) - \frac{i}{2} \varepsilon^{\alpha \beta \gamma \delta} Tr ({\bar Z}_{\delta} , [{\bar {\psi}}_{\alpha} , \psi_{\beta} ; {\bar Z}_{\gamma} ]),
\end{eqnarray}
with
\begin{equation}
V = \frac{2}{3} Tr (\Upsilon^{\gamma \delta}_{\beta} , {\bar{\Upsilon}}^{\beta}_{\gamma \delta}),
\end{equation}
and
\begin{equation}
\Upsilon^{\gamma \delta}_{\beta}= [Z^{\gamma}, Z^{\delta} ; {\bar Z}_{\beta} ] - \frac{1}{2} \delta^{\gamma}_{\beta} [Z^{\alpha}, Z^{\delta} ; {\bar Z}_{\alpha}] + \frac{1}{2} \delta^{\delta}_{\beta} [Z^{\alpha}, Z^{\gamma}; {\bar Z}_{\alpha} ].
\end{equation}
Supersymmetry transformations are redressed as  in Ref. \cite{Passirini,Low} and  the supercurrent  is given  as follows:
\begin{equation}
J_\mu =  {\bar{\varepsilon}}^I J^I_\mu = Tr (\delta {\bar \psi}_{\alpha}  \gamma_\mu , \psi^{\alpha} ) + Tr (\delta {\bar \psi}^{\alpha} \gamma_\mu , \psi_{\alpha}).\label{supercurrent}
\end{equation}
Boundary conditions of $N=6$ BLG theory \cite{Aali1}  was investigated in the same method as for boundary conditions of $N=8$ BLG theory \cite{Aali1} was done. The attainment of  vanishing the normal component of supercurrent in the boundary conditions is following relation:
\begin{eqnarray}
\nonumber
0 &=&  \Gamma^I_{\alpha \beta} {\bar{\varepsilon}}^I \gamma^\mu D_\mu Z^{\beta} \gamma_2 \psi^{\alpha} -  \Gamma^I_{\alpha \beta} [Z^{\gamma} , Z^{\beta}; {\bar Z}_{\gamma} ] {\bar{\varepsilon}}^I  \gamma_2 \psi^{\alpha} - \Gamma^I_{\gamma \delta} [Z^{\gamma}, Z^{\delta} ; {\bar Z}_{\alpha} ] {\bar{\varepsilon}}^I  \gamma_2 \psi^{\alpha}
\\
&-& {\tilde \Gamma}^{I\alpha \beta} {\bar{\varepsilon}}^I \gamma^\mu D_\mu {\bar Z}_{\beta} \gamma_2  \psi_{\alpha} - {\tilde\Gamma}^{I\alpha \beta} [{\bar Z}_{\gamma} , {\bar Z}_{\beta} ; Z^{\gamma}] \varepsilon^I \gamma_2  \psi_{\alpha}-  {\tilde\Gamma}^{I\gamma \delta} [{\bar Z}_{\gamma},{\bar Z}_{\delta} ; Z^{\alpha}] \varepsilon^I \gamma_2  \psi_{\alpha}.
\end{eqnarray}
 As we told the method is the same method in \cite{Aali1} with some difference. One of them is decomposition of $SU(4)$ as Lorentzian symmetry into two groups. in this case, one way is  decompose 4 complex scalar field into $X^{\alpha}={Z^1,Z^2}$ and $Y^{{\alpha}^{'}}={Z^2,Z^3}$ as $SO(1,1) \times SU(2) \times SU(2) \subset SO(1,2) \times SU(4)$ \cite{Sezgin}:
\begin{eqnarray}
\nonumber
0 &=&   \Gamma^I_{\alpha \beta} {\bar{\varepsilon}}^I \gamma^{\hat\mu} D_{\hat\mu} X^{\beta} \gamma_2 \psi^{\alpha} \\
\nonumber
&-& {\tilde \Gamma}^{I{\alpha}^{'}\beta} {\bar{\varepsilon}}^I \gamma^{\hat\mu} D_{\hat\mu} {\bar X}_{\beta} \gamma_2  \psi_{{\alpha}^{'}}
\\
\nonumber
&+&   \Gamma^I_{{\alpha}^{'}{\beta}^{'}} {\bar{\varepsilon}}^I \gamma^{\hat\mu} D_{\hat\mu} Y^{{\beta}^{'}} \gamma_2 \psi^{{\alpha}^{'}} \\
\nonumber
&-& {\tilde \Gamma}^{I\alpha {\beta}^{'}} {\bar{\varepsilon}}^I \gamma^{\hat\mu} D_{\hat\mu} {\bar Y}_{{\beta}^{'}} \gamma_2  \psi_{\alpha}
\\
\nonumber
&+& \Gamma^I_{\alpha \beta} {\bar{\varepsilon}}^I \gamma^2 D_2 X^{\beta} \gamma_2 \psi^{\alpha} -  \Gamma^I_{\alpha \beta} [X^{\gamma} , X^{\beta}; {\bar X}_{\gamma} ] {\bar{\varepsilon}}^I  \gamma_2 \psi^{\alpha}- \Gamma^I_{\gamma \delta} [X^{\gamma} , X^{\delta} ; {\bar X}_{\alpha} ] {\bar{\varepsilon}}^I  \gamma_2 \psi^{\alpha}
\\
\nonumber
&+& \Gamma^I_{{\alpha}^{'}\beta} {\bar{\varepsilon}}^I \gamma^2 D_2 X^{\beta} \gamma_2 \psi^{{\alpha}^{'}} -  \Gamma^I_{{\alpha}^{'}\beta} [X^{\gamma}, X^{\beta} ; {\bar X}_{\gamma} ] {\bar{\varepsilon}}^I  \gamma_2 \psi^{{\alpha}^{'}}
\\
\nonumber
&+& \Gamma^I_{{\alpha}^{'}{\beta}^{'}} {\bar{\varepsilon}}^I \gamma^2 D_2 Y^{{\beta}^{'}} \gamma_2 \psi^{{\alpha}^{'}} -  \Gamma^I_{{\alpha}^{'}{\beta}^{'}} [Y^{{\gamma}^{'}} , Y^{{\beta}^{'}} ; {\bar Y}_{{\gamma}^{'}} ] {\bar{\varepsilon}}^I  \gamma_2 \psi^P - \Gamma^I_{{\beta}^{'}{\gamma}^{'}} [Y^{{\beta}^{'}}, Y^{{\gamma}^{'}} ; {\bar Y}_{{\alpha}^{'}} ] {\bar{\varepsilon}}^I  \gamma_2 \psi^{{\alpha}^{'}}
\\
\nonumber
&+& \Gamma^I_{\alpha {\beta}^{'}} {\bar{\varepsilon}}^I \gamma^2 D_2 Y^{{\beta}^{'}} \gamma_2 \psi^{\alpha}-  \Gamma^I_{\alpha {\beta}^{'}} [Y^{{\gamma}^{'}} , Y^{{\beta}^{'}} ; {\bar Y}_ {{\gamma}^{'}}] {\bar{\varepsilon}}^I  \gamma_2 \psi^{\alpha}
\\
\nonumber
&-& \Gamma^I_{{\beta}^{'}\beta} [Y^{{\alpha}^{'}} , X^{\beta} ; {\bar Y}_{{\alpha}^{'}} ] {\bar{\varepsilon}}^I  \gamma_2 \psi^{{\beta}^{'}} - \Gamma^I_{\gamma {\beta}^{'}} [X^{\gamma} , Y^{{\beta}^{'}} ; {\bar Y}_{{\alpha}^{'}} ] {\bar{\varepsilon}}^I  \gamma_2 \psi^{{\alpha}^{'}}- \Gamma^I_{\alpha \beta} [Y^{{\alpha}^{'}} , X^{\beta} ; {\bar Y}_{{\alpha}^{'}} ] {\bar{\varepsilon}}^I  \gamma_2 \psi^{\alpha}
\\
\nonumber
&-& \Gamma^I_{{\alpha}^{'}{\beta}^{'}} [X^{beta} , Y^{{\alpha}^{'}} ; {\bar X}_{\beta}] {\bar{\varepsilon}}^I  \gamma_2 \psi^{{\beta}^{'}} -  \Gamma^I_{\alpha {\alpha}^{'}} [X^{\gamma}, Y^{{\alpha}^{'}} ; {\bar X}_{\gamma} ] {\bar{\varepsilon}}^I  \gamma_2 \psi^{\alpha}-\Gamma^I_{{\alpha}^{'}{\beta}^{'}} [Y^{{\alpha}^{'}} , Y^{{\beta}^{'}} ; {\bar X}_{\alpha}] {\bar{\varepsilon}}^I  \gamma_2 \psi^{\alpha}
\\
\nonumber
&-& \Gamma^I_{\gamma \delta} [X^{\gamma} , X^{delta} ; {\bar Y}_{{\alpha}^{'}} ] {\bar{\varepsilon}}^I  \gamma_2 \psi^{{\alpha}^{'}}
\\
\nonumber
&-&\Gamma^I_{\gamma {\beta}^{'}} [X^{\gamma} , Y^{{\beta}^{'}} ; {\bar X}_{\alpha} ] {\bar{\varepsilon}}^I  \gamma_2 \psi^{\alpha}
\\
\nonumber
&-& {\tilde\Gamma}^{I{\beta}^{'}{\alpha}^{'}} [{\bar X}_{\gamma} , {\bar Y}_{{\alpha}^{'}} ;X^{\gamma}] \varepsilon^I \gamma_2  \psi_{{\beta}^{'}} 
\\
\label{90}
&-& {\tilde\Gamma}^{I{\alpha}^{'}\beta} [{\bar Y}_{{\alpha}^{'}}, {\bar X}_{\beta} ;Y^{{\alpha}^{'}}] \varepsilon^I \gamma_2  \psi_{{\alpha}^{'}} 
,
\end{eqnarray}
where $\tilde \Gamma_2\psi_a=\tilde\psi_a$.  Different Lorentzian symmetry for each has been considered,   individually, then one can take zero each one with different Lorentzian symmetry and solve. In order to this purpose, the different type of boundary conditions can be provided which we have investigated only the Dirichlet boundary conditions in half of the scalar fields, i.e.,  $D_{\hat\mu}Y=0$ and simplest solution for it  i.e. $Y=0$,  maintain equations of (\ref{90}) are as follows:
\begin{eqnarray}
0 &=&   \Gamma^I_{\alpha \beta} {\bar{\varepsilon}}^I \gamma^{\hat\mu} D_{\hat\mu} X^{\beta} \gamma_2 \psi^{\alpha}
\\
0&=& {\tilde \Gamma}^{I{\alpha}^{'}\beta} {\bar{\varepsilon}}^I \gamma^{\hat\mu} D_{\hat\mu} {\bar X}_{\beta} \gamma_2  \psi_{{\alpha}^{'}}
\\
0 &=&   \Gamma^I_{{\alpha}^{'}{\beta}^{'}} {\bar{\varepsilon}}^I \gamma^{\hat\mu} D_{\hat\mu} Y^{{\beta}^{'}} \gamma_2 \psi^{{\alpha}^{'}} \\
0&=& {\tilde \Gamma}^{I\alpha {\beta}^{'}} {\bar{\varepsilon}}^I \gamma^{\hat\mu} D_{\hat\mu} {\bar Y}_{{\beta}^{'}} \gamma_2  \psi_{\alpha}
\\
0&=& \Gamma^I_{\alpha \beta} {\bar{\varepsilon}}^I \gamma^2 D_2 X^{\beta} \gamma_2 \psi^{\alpha}-  \Gamma^I_{\alpha \beta} [X^{\gamma} , X^{\beta} ; {\bar X}_{\gamma} ] {\bar{\varepsilon}}^I  \gamma_2 \psi^{\alpha} - \Gamma^I_{\gamma \delta} [X^{\gamma} , X^{\delta} ; {\bar X}_{\alpha} ] {\bar{\varepsilon}}^I  \gamma_2 \psi^{\alpha}
\\
\label{BASU}
0&=& \Gamma^I_{{\alpha}^{'}\beta} {\bar{\varepsilon}}^I \gamma^2 D_2 X^{\beta} \gamma_2 \psi^{{\alpha}^{'}} -  \Gamma^I_{{\alpha}^{'}\beta} [X^{\gamma}, X^{\beta} ; {\bar X}_{\gamma} ] {\bar{\varepsilon}}^I  \gamma_2 \psi^{{\alpha}^{'}}
.
\end{eqnarray}
Introducing projection operator for solving these equations is the main work now \cite{Sezgin}. Relation (\ref{BASU}) is the only condition independent of $\Gamma$ on scalar fields, which  is Basu-Harvey equation with  $F^{ABC}\hspace{0cm}_D$ as the structure constant of the  Manin triple of  3-Leibniz algebra $\cal D$ :
\begin{eqnarray}
0 = D_2X^I_A + \frac{1}{6} \epsilon^{IJKL} X^J_B X^K_C X^L_D  F^{BCD}_{\phantom{BCD}A}.
\end{eqnarray}

In the previous sections, we introduced a special example of 3-Leibniz bialgebra in relation to Lie bialgebra. The result of applying that example  and using commutation relations mentioned in  (\ref{123456}) is the Basu-Harvey equation with the following form:
\begin{eqnarray}
\label{2}
0 &=& D_2X^I_- + \frac{1}{6} \epsilon^{IJKL} X^J_b X^K_c X^L_d  F^{BCD}_{\phantom{abc}-} , 
\\
\label{3}
0 &=& D_2X^I_{\tilde -} + \frac{1}{6} \epsilon^{IJKL} X^J_B X^K_CX^L_D  F^{BCD}_{\phantom{abc}\tilde -} , 
\\
\nonumber
 0&=& D_2X^I_i + \frac{1}{6} \epsilon^{IJKL} X^J_B X^K_C X^L_D  F^{BCD}_{\phantom{abc}i} \\
\label{4}
 &=& D_2X^I_i + \frac{1}{2} g_{YM}\epsilon^{IJK}  X^J_j X^K_k  f^{jk}_{\phantom{abc}i} + \frac{1}{2} g_{YM}\epsilon^{ACD}  X^{J\tilde j} X^K_k  {\tilde f}_{ij}\hspace{0cm}^{k}, 
\\
\nonumber
0 &=& D_2X^I_{\tilde i} + \frac{1}{6} \epsilon^{IJKL} X^J_B X^K_C X^L_D  F^{BCD}_{\phantom{abc}\tilde i} \\
\label{5}
&=& D_2X^I_{\tilde i}+ \frac{1}{2} g_{YM}\epsilon^{IKL}  X^{K\tilde j} X^L_{\tilde k}  {\tilde f}_{jk}^{\phantom{abc}\tilde i} + \frac{1}{2} g_{YM}\epsilon^{IKL}  X^{K\tilde j} X^L_k  {\tilde f}_{\tilde ij}\hspace{0cm}^k.
\end{eqnarray}
vanishing equations (\ref{2},\ref{3}) is a result of existence  of two totally antisymmetric coefficients with ($I,J,K,L$) and  ($A,B,C=-,+,i,{\tilde -},{\tilde +}, {\tilde i}$ and $i, j, k$) index ($A,B,C=-,+,i,{\tilde -},{\tilde +}, {\tilde i}$ and $i, j, k$), i.e., $ \epsilon^{IJKL}$ and $ F^{BCD}_{\phantom{abc}-}$. Equations (\ref{4},\ref{5}) can be combined and written in following form:
\begin{eqnarray}
\partial_\sigma X^I _{\bar A}&=&   
\frac{1}{2} \epsilon_{JKI}X^J_{\bar B} X^K_{\bar C} F^{{\bar B}{\bar C}}\hspace{0cm}_{\bar A},
\end{eqnarray} 
 where $ F^{{\bar B}{\bar C}}\hspace{0cm}_{\bar A}$ is the structure constant for ${\cal D}_{\cal G}$ with  ${\bar A},{\bar B},{\bar C}=i,{\tilde i}$ which is the  Manin triple of  Lie algebra $\cal G$. It has been considered as  Nahm equation a result of considering boundary conditions in D1-branes\cite{chu}  with the Lie bialgebra representation.
  So, one of the boundary conditions of the BL model for multiple membranes is Basu-Harvey equation is in relation to Nahm equation  \cite{chu} and vice versa as BLG one in  Ref. \cite{Aali1} for 3-Lie bialgebra\footnote{Relation between Nahm and Basu- Harvey has mentioned in Ref. \cite{Medeiros} by using of Lie superalgebra \cite{Medeiros}.}. According to the
obtained relation between Nahm and Basu-Harvey equations, there is the relation between M2 and D2 actions and
vice versa. As you know, Nahm equation is a result in the studying boundary condition for DBI action\cite{chu} and according to the relation obtained between Nahm and Basu-Harvey equations, it could manifest relation between M2 and D2 actions and vice versa. This show relation between M-theory and string theory and vice versa because BL model is a model for membranes which have been derived from supergravity action as low energy limit of M-theory and DBI is the similar one in the string theory.
\section*{ Conclusions}
 We have constructed  $N=6$ Chern-Simons gauge theory  (BL model) on a special  Manin triple  $({\cal D},{\cal A},{\cal A}^*)$ as a 3-Leibniz bialgebra. Then using the correspondence   between 3-Leibniz bialgebra $({\cal A},{\cal A}^*)$ and Lie bialgebra $({\cal G},{\cal G}^*)$, we have shown a relation between $N=6$  BL model  with 3-Leibniz algebra structure and  $N=(4,4)$ WZW model over Lie group with Manin triple (as Lie algebra) of  Lie  bialgebra  structure. Also,   boundary conditions for BL model on Manin triple were considered and the Basu-Harvey equation reduced to Nahm equation, in this way.
\\
Conversely; if  $N = (4,4)$ supersymmetric WZW
action with a Lie bialgebra equipped one 2-cocycle is constructed by helping DBI-action for D2-brane, then constructing M2-model on the special 3-Leibniz algebra is possible according to the relation between Lie bialgebra
and 3-Leibniz bialgebra. The other result of  this correspondence is obtaining Basu-Harvey equation by considering Nahm equation as the boundary conditions of the D1-string ending to D3-brane in the following relation with DBI action.

\section*{ Acknowledgments}
We would like to thank  M. Akbari-Moghanjoughi  for carefully reading the manuscript. This research was supported by a research fund
No. 217D4310 of Azarbaijan Shahid Madani university.

\end{document}